\documentclass{PoS}
\usepackage{graphicx}
\usepackage{amssymb}
\usepackage{epstopdf}
\usepackage{layout}
\usepackage{subfigure}

\title{Gamma-Gamma Absorption in Gamma-Ray Burst Environments}

\ShortTitle{Gamma Gamma Absorption in GRB Environments}

\author{\speaker{Lent\'e Dreyer}\\
        Centre for Space Research, North-West University, Potchefstroom, 2520, South Africa\\
        E-mail: \email{lentedreyer@gmail.com}}  

        \author{Markus B\"ottcher\thanks{The work of M. B. is supported through the South African
Research Chair Initiative (SARChI) of the South African National Research Foundation (NRF) and the 
Department of Science and Technology, under SARChI Chair grant no. 64789. {\it Disclaimer:} Any 
opinion, finding and conclusion or recommendation expressed in this material is that of the authors 
and the NRF does not accept any liability in this regard. }\\
        Centre for Space Research, North-West University, Potchefstroom, 2520, South Africa\\
        E-mail: \email{Markus.Bottcher@nwu.ac.za}}        

\abstract{Gamma-ray bursts (GRBs) are the most violent explosions in the universe, seen primarily as bright, 
short flashes of $\gamma$-rays. Long GRBs are most likely associated with the violent death of a very massive star. 
They are thus believed to originate within regions of recent or ongoing star formation with various bright, 
young stars, for instance OB associations. GRBs have been detected in $\gamma$-rays by satellite-borne 
$\gamma$-ray telescopes at MeV -- GeV energies, but could potentially also emit $\gamma$-rays at even higher 
energies (Very-High-Energy [VHE] $\gamma$-rays: $E > 100$~GeV), as observed by ground-based Cherenkov Telescope 
facilities, such as the High Energy Stereoscopic System (H.E.S.S) in Namibia. VHE gamma-rays can be absorbed 
by low energy photons by pair production, and the stars in the vicinity of a GRB provide a dense radiation 
environment, which could lead to such absorption. We have investigated representative GRB environments and 
provide an estimate of the opacity to $\gamma\gamma$ absorption for VHE $\gamma$-rays from GRBs. We find 
that for the likely properties of OB associations around GRB progenitors, $\gamma\gamma$ absorption is 
expected to be negligible.}

\FullConference{4th Annual Conference on High Energy Astrophysics in Southern Africa ,\\
		25 - 26 August 2016\\
		South African Astronomical Observatory (SAAO), Cape Town, South Africa}

\begin{document}

\section{\label{intro}Introduction}
Gamma-ray bursts (GRBs) are the most violent explosions in the universe, seen primarily as bright, short flashes 
of $\gamma$-rays. GRBs have been detected in $\gamma$-rays up to a few GeV thus far (e.g., \cite{Abdo2009}), 
and although they have not been detected at higher energies by ground-based Cherenkov Telescopes yet, considerable 
progress has been made with developments in observations, theory, and instrumentation in order to make this 
possible \cite{Williams2009}. GRBs are randomly distributed in the sky \cite{Fishman1995}, and there are 
no repeating GRB sources \cite{RB1994}. Most GRBs have gradually decaying afterglows in $X$-rays, some also in 
optical and radio and even extended (in time) emission in high-energy ($E > 100$~MeV) $\gamma$-rays (e.g., \cite{Hurley1994}). 
There are at least two classes of GRBs, referred to as long- and short GRBs, depending on their duration, $t_{90}$ 
(the time between 5~\% and 95~\% of the gamma-ray fluence being received), with short GRBs having $t_{90} \lesssim 2$~s 
and long GRBs having $t_{90} \gtrsim 2$~s, up to several minutes. These two classes also differ in the spectral 
hardness of their sub-$MeV$ emission, with short GRBs typically having harder spectra (e.g., \cite{Dezalay1996}). 
The observed brightness, combined with the large distances, implies isotropic-equivalent energy outputs of 
$E_{\rm iso} \sim 10^{51}$~erg (e.g., \cite{Fishman1995}). 

Although the exact origin of GRBs is still under investigation, the two classes are generally believed to have different 
progenitors based on their host environments. Long GRBs, for instance, are often found in spiral arms (star forming 
regions) of very faint host galaxies, while short GRBs are generally found far from star-forming regions of
their host galaxies. The leading model for long GRBs involves a supernova-like explosion of a very 
massive star ($> 25 \, M_{\odot}$), where the core collapse leads to the direct formation of a black hole. The rest 
of the star is accreted onto the newly-formed black hole, resulting in the formation of an accretion disk which, in 
turn, powers ultra-relativistic jets. These jets seem to be the source of the observed $\gamma$-rays 
\cite{McFadyen1999,Levan2016}. 

Short GRBs are likely caused by the coalescence of a black hole and a neutron star or a neutron-star binary.  
The neutron star will be destroyed by tidal disruption, and its matter will accrete onto the (potentially newly-formed
in the process) black hole. This, as in the case of long GRBs, leads to the formation of an accretion disk and the 
explosion of ultrarelativistic jets. Such events are also expected to produce strong gravitational-wave signals,
potentially detectable by LIGO/VIRGO (e.g., \cite{GS09}). 

The very detection of VHE $\gamma$-rays emission from GRBs would severely constrain the physical properties 
pertaining to the particle acceleration by GRBs, and would therefore severely constrain models for particle 
acceleration and the total energy budget, and hence the properties of the GRB progenitors 
\cite{Hurley1995,Aharonian2004,Williams2009}.
Detections of GRBs at MeV -- GeV energies is now routinely provided by the Fermi Gamma-Ray Space Telescope 
(e.g., \cite{Abdo2009}), and all of the current Imaging Cherenkov Telescopes have active observing programmes
towards the detection of VHE $\gamma$-ray emission from GRBs (e.g., \cite{Aharonian09}). 

Very High Energy (VHE) $\gamma$-rays are subject to $\gamma\gamma$ pair production of an electron-positron pair 
through the interaction with optical/infrared photons \cite{Gould1967}. Hence, in addition to the question whether
particles are accelerated to sufficient energy to produce VHE $\gamma$-ray emission, it is also important to
understand their chance of propagating to Earth unabsorbed. An important target photon field for VHE $\gamma$-ray
absorption is the Extragalactic Background Light (EBL), which is generally believed to limit the distance to which
VHE $\gamma$-ray sources can realistically be detected by ground-based Atmospheric Cherenkov Telescope facilities,
to $z \lesssim 1$ (e.g., \cite{Franceschini08,Finke10}). $\gamma\gamma$ absorption may also play an important role
within the $\gamma$-ray emission region of a GRB itself and is likely to have a profound impact on the shape of the 
emerging $\gamma$-ray spectrum (e.g., \cite{Pe2004}. However, also photon fields in the local environment of the 
source, within the host galaxy, may potentially be relevant and affect the detectability of VHE emission from 
GRBs. 

The purpose of this work is to investigate representative GRB environments and provide an estimate of the opacity 
to $\gamma\gamma$-absorption that VHE $\gamma$-rays from the GRB would be subject to. Short GRBs are known to 
originate in rather ``clean'' environments, typically far from bright light sources (such as star-forming regions),
so that no $\gamma\gamma$ absorption by local radiation fields is expected to be relevant. However, long GRBs,
resulting from the deaths of very massive (and thus, young) stars, are expected to occur within young star-forming
regions (O/B associations), and these are the environments which we will focus on in this study. We therefore
consider typical O/B associations likely to host GRB progenitors, and calculate the $\gamma\gamma$-opacity for 
VHE $\gamma$-rays produced by the GRB in such an environment.

\section{\label{GGabsorption}$\gamma\gamma$-Absorption and Pair Production}

High-energy $\gamma$-ray photons can interact with each other, and with lower energy photons, to produce an 
electron-positron pair. This is the process of $\gamma\gamma$-absorption, which is the inverse process of 
annihilation and is one of the most relevant elementary processes in high-energy astrophysics \cite{Aharonian2004}. 
The role of pair production was first pointed out by Gerasimova \cite{Gerasimova1962}, and emphasized by 
Bonometto and Rees in the context of dense radiation fields of very compact objects in 1971 \cite{Bonometto1971}. 
When a photon of dimensionless energy $\varepsilon_1 = E / m_ec^2$ interacts with a photon of energy 
$\varepsilon_2$, with an interaction angle $\theta$, the resulting electron-positron pair will, in the 
center-of-momentum ($cm$) frame, have equal Lorentz factors, $\gamma_{cm} = \sqrt{\varepsilon_1 \varepsilon_2 
(1- \mu) / 2}$ with $\mu = \cos\theta$. The normalized velocities will also be the same (but in the opposite 
direction) and are given by  
\begin{eqnarray}
\label{eq:velocities}
\beta_{cm} = \sqrt[]{1- \frac{2}{\varepsilon_1 \varepsilon_2 (1- \mu)}}
\end{eqnarray}
The threshold condition for $\gamma\gamma$ absorption results from the requirement that $\gamma_{cm} \geq 1$,
which provides the threshold condition for $\varepsilon_1$,
\begin{eqnarray}
\label{eq:threshold}
\varepsilon_1 \geq \frac{1}{\varepsilon_2(1-\mu)}
\end{eqnarray}
The most favorable interaction angle corresponds to head-on collision, where $\mu =-1$, so that the absolute 
threshold energy is $\varepsilon_{1, \rm thr} = 1 / \varepsilon_2$. The full $\gamma \gamma$ cross section 
is given by \cite{Gould1967}
\begin{eqnarray}
\label{eq:fullcrosssection}
\sigma_{\gamma\gamma} = \frac{3}{16} \sigma_T (1 - \beta_{cm}^2) \times \left( [3-\beta_{cm}^4] 
\ln \left( \frac{1 + \beta_{cm}}{1 - \beta_{cm}} \right) - 2\beta_{cm} [2-\beta_{cm}^2] \right)
\end{eqnarray}
where $\sigma_T$ is the Thompson cross-section, defined as 
\begin{eqnarray}
\sigma_T = \frac{8\pi}{3} \, r_0^2
\end{eqnarray}
The energy dependence of $\sigma_{\gamma\gamma}$ is strongly 
peaked around $\varepsilon_1 = 2 \varepsilon_{1, \rm thr}$, with the threshold energy,
\begin{eqnarray}
\varepsilon_{th} = \frac{1}{\varepsilon_1(1-\mu)}
\end{eqnarray}
When the energy of the $\gamma-$ray is exactly equal to $\varepsilon_{1, \rm thr}$, 
the normalized velocities, as well as $\sigma_{\gamma\gamma}$ itself, will be zero (e.g., \cite{Boettcher2012}). 

\begin{figure}[ht]
\vspace{0.3cm}
\centering
\includegraphics[width=10cm]{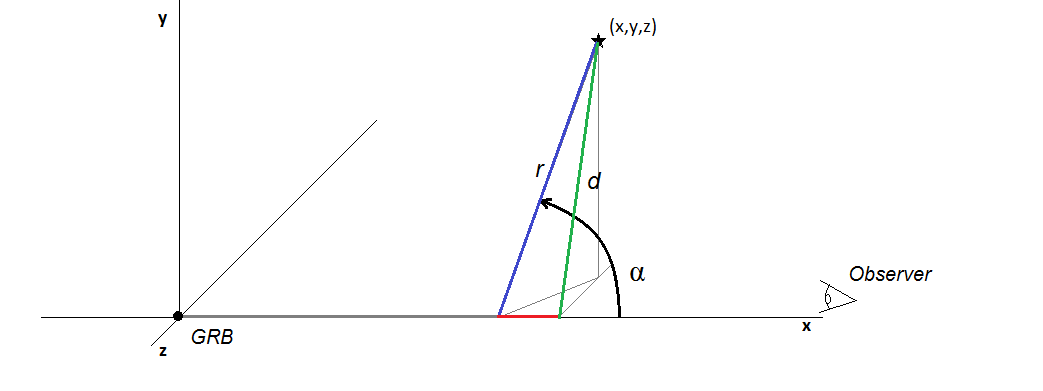}
\caption{\label{fig:Geometry}The geometry for the consideration of a single star. The GRB is at the origin, 
with the observer along the line of sight ($x$-axis). The distance $d$ is the shortest distance between the
line of sight and the star, while $r$ is the distance between the star and any given point $x$ along the line
of sight. }
\end{figure}

The opacity $\tau_{\gamma\gamma}(\varepsilon_1)$ for $\gamma\gamma$ absorption of a $\gamma$ ray photon, with energy $\varepsilon_1$, in a photon field $n_{ph}(\varepsilon_2,x)$ is given, over a path length $\ell$, by
\begin{eqnarray}
\label{eq:gammaopacity1}
\tau_{\gamma\gamma}(\varepsilon_1) = \int^\ell_0 dx \int^1_{-1} (1-\mu) d\Omega \int^\infty_{\frac{1}{\varepsilon_1(1-\mu)}}  n_{ph} (\epsilon_2, x) \sigma_{\gamma\gamma}(\varepsilon_1, \varepsilon_2, \mu) d\varepsilon_2
\end{eqnarray}
where $x$ is the distance from the GRB along the $\gamma$-ray path and the solid angle element $d\Omega$ is $d\Omega = d\phi d\mu$ (see Fig. \ref{fig:Geometry}).
If the absorbing radiation field is located outside the emission region, $\gamma\gamma$-absorption leads to an 
exponential factor compared to the intrinsic flux, produced in the emission zone, i.e.,
\begin{eqnarray}
F_\nu^{obs}(\varepsilon) = F_\nu^{int}(\varepsilon) e^{-\tau_{\gamma\gamma}(\varepsilon)}
\end{eqnarray}
where $F_\nu^{obs}(\varepsilon)$ is the observed flux, and $F_\nu^{int}(\varepsilon)$ is the intrinsic flux.

\section{\label{Model}Model Setup and Calculations}

We perform our calculation in a co-ordinate system in which the GRB occurs at the origin, with the observer along 
the $x$-axis. Stars of the O/B association hosting the GRB progenitor are distributed randomly in a volume characteristic
for such clusters around the GRB. The geometry for the consideration of a single star located at co-ordinates
$(x_{\ast}, y_{\ast}, z_{\ast})$ is illustrated in figure 
\ref{fig:Geometry}. The interaction angle $\theta$ is defined as $\pi - \alpha$, where $\alpha = \cos^{-1} (d/r)$, 
with $d$ being the shortest distance between the star and the line of sight (i.e., the impact parameter), and $r$ 
the current distance between the star and the point $x$ along the line of sight, 
\begin{eqnarray}
d = \sqrt[]{y_{\ast}^2 +z_{\ast}^2} \\
r = \sqrt[]{(x_{\ast} - x)^2 + y_{\ast}^2 +z_{\ast}^2}
\end{eqnarray}
OB-associations are loosely organized, very young groups of stars that formed approximately at the same time, 
containing a significant fraction of O and B stars (e.g., \cite{Reed2003}). They commonly contain between $10$ 
and $100$ O and B stars, with a total mass between hundreds and thousands of solar masses, and radii between $\sim 8$ 
and $160$ parsec (e.g., \cite{Lang2012}). The stars each have a corresponding luminosity and temperature, taken 
from \cite{Carroll1996}. Compared to the dimensions of the O/B association, the stars are small enough to be 
considered as point sources. This simplifies the calculation of the $\gamma\gamma$ opacity according to equation 
\ref{eq:gammaopacity1}, as we can substitute the integrals over $d\Omega = d\phi \, d\mu$ with a summation over 
all stars, 
\begin{eqnarray}
\label{eq:gammaopacity2}
\tau_{\gamma\gamma}(\varepsilon_1) = \int_0^\ell dx \, \sum^N_{k=1} (1 - \mu_k) 
\int\limits^\infty_{\frac{1}{\varepsilon_1 (1 - \mu_k)}} n_{ph, k} (\varepsilon_2, x) 
\sigma_{\gamma\gamma} (\varepsilon_1, \varepsilon_2, \mu_k) d\varepsilon_2
\end{eqnarray}
for a total number of $N$ stars that are taken into consideration. $n_{ph, k}$ is the density spectrum of photons
from star $k$ at point $x$ along the line of sight, with which the $\gamma$-ray interacts under an interaction-angle 
cosine $\mu_k$ (as a function of $x$). 

The spectra of the O and B stars are approximated as blackbody spectra, corresponding to their temperatures $T$, and 
luminosities $L$. Proper normalization to the stellar luminosity yields for the photon number density spectrum of an
individual star:
\begin{eqnarray}
n_{ph}(\varepsilon) = {15 \, L \over \pi^2 \Theta^4} \, \frac{\varepsilon^2}{(e^{\varepsilon/\Theta}-1) 
4\pi r^2\varepsilon m_ec^3}
\end{eqnarray}
where $\Theta = k T / (m_e c^2)$ is the dimensionless temperature of the star.

\begin{figure}[ht]
\vspace{0.3cm}
\centering
\includegraphics[width=10cm]{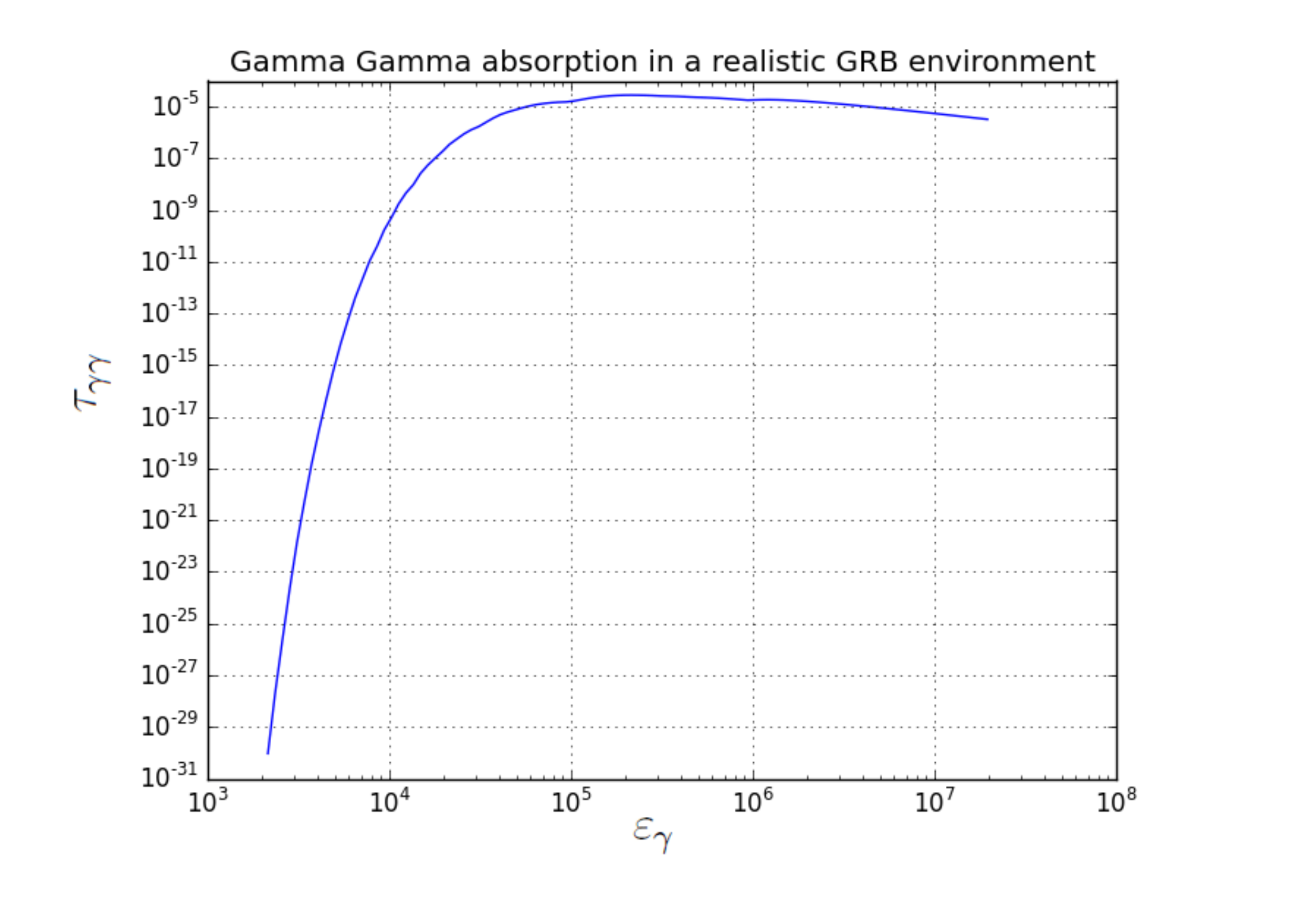}
\caption{\label{fig:Result}The $\gamma\gamma$ opacity, $\tau_{\gamma\gamma}$, as a function of the $\gamma$-ray energy, $\varepsilon_{\gamma}$ (in units of $m_e c^2$), 
in a representative realization of the distribution of 60 O/B stars within a radius of $100$ parsec.}
\end{figure}

\section{\label{Results}Results and Discussion}

As a representative characterization of a typical GRB environment, we consider in our calculations $60$ O/B stars with 
a typical Salpeter initial mass function.  O and B stars are characterized by masses of $ 3.8 M_{\odot} \lesssim M \lesssim 60 M_{\odot}$ and
    luminosities of $ 60 L_{\odot} \lesssim L \lesssim 49900 L_{\odot}$ that form in loosely organized groups called OB associations \cite{Carroll1996}.
We consider these stars to be randomly evenly distributed within a radius of $100$ parsec around 
the GRB source. The $\gamma\gamma$-opacity has been calculated in an energy range between $1$~GeV and $10$~TeV, as the
opacity at lower energies is found to be negligibly small due to the characteristic optical / UV range of the stellar 
spectra. The $\gamma\gamma$ opacity as a function of $\gamma$-ray energy, is shown in figure \ref{fig:Result}, for one
representative realization of the stellar distribution. 

The $\gamma\gamma$ opacity peaks in the $\gamma$-ray energy range of $20$~GeV and $100$~GeV, and slowly decreases
towards higher energies due to the decrease of the $\gamma\gamma$ cross section with increasing $cm$ energy in
interactions of optical/UV photons and the decrease of the photon number densities at energies $\sim 2/\varepsilon$
with increasing $\varepsilon$ (i.e., decreasing target photon energy towards the IR). The maximum $\gamma\gamma$ 
opacity in this specific realization is of order $10^{-5}$. We performed the same calculation for a large number 
of realizations of the stellar distribution and found similar results in each case. 

\begin{figure}[ht]
\vspace{0.3cm}
\centering
\includegraphics[width=15cm]{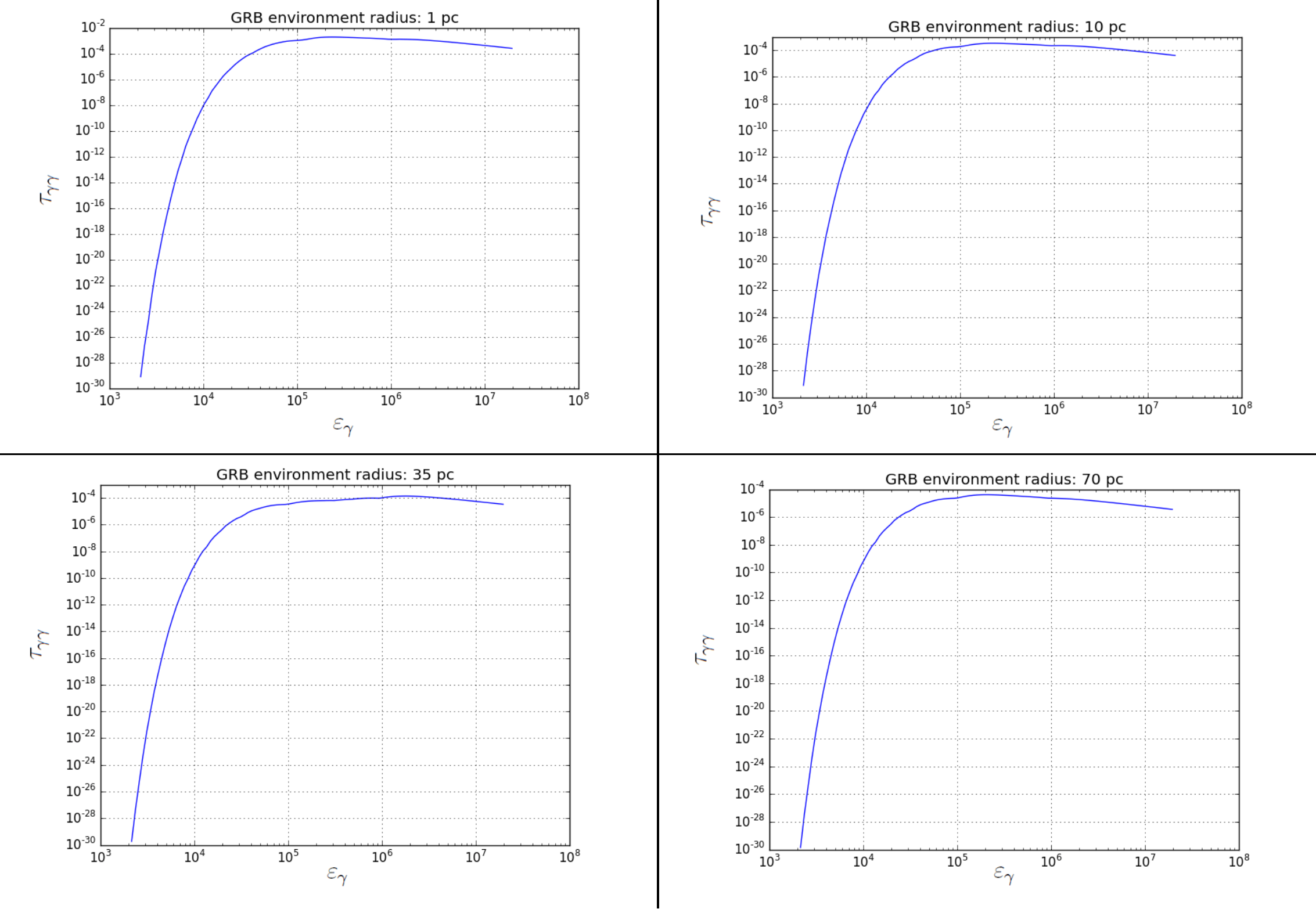}
\caption{\label{fig:ParameterStudy} A parameter study of the dependence of the $\gamma\gamma$ opacity on the 
radius of the O/B association containing 60 O/B stars. $\tau_{\gamma\gamma}$ is plotted as a function of
$\varepsilon_{\gamma}$ for O/B association radii from $1$ parsec (top left), to $75$ parsec (bottom right).}
\end{figure}

In order to investigate the sensitivity of our result on the specifics of the stellar distribution, we conducted 
a parameter study, in which we varied the radius of the O/B association between $1$ parsec and $100$ parsec, containing
the same number of $60$ O/B stars. Characteristic results are shown in figure \ref{fig:ParameterStudy}, where again the 
$\gamma\gamma$ opacity is plotted as a function of the $\gamma$-ray energy in the energy range of $\sim 1$~GeV to $10$~TeV. 
As expected, the $\gamma\gamma$ opacity increases with decreasing radius of the O/B association (i.e., increasing density 
of stars), but even in the extreme (unrealistic) case of 60 O/B stars within 1 parsec, the opacity still does not exceed
$\sim 10^{-2}$. 

We therefore conclude that the $\gamma\gamma$ opacity due to an O/B association hosting a GRB progenitor is expected
to be negligible for any realistic configuration. VHE $\gamma$-rays are, therefore, highly unlikely to be significantly
attenuated by the photon fields of their host O/B associations.

\end{document}